# Positron acceleration in a hollow plasma channel up to TeV regime

**Longqing Yi**[1], **Baifei Shen**[1], **Liangliang Ji**[1], **Konstantin Lotov**[2,3], **Alexander Sosedkin**[2,3], **XiaomeiZhang**[1], **Wenpeng Wang**[1], **Jiancai Xu**[1], **Yin Shi**[1], **Lingang Zhang**[1] **& Zhizhan Xu**[1]

[1]State Key Laboratory of High Field Laser Physics, Shanghai Institute of Optics and Fine Mechanics, Chinese Academy of Sciences, P.O. Box 800-211, Shanghai 201800, China, [2]Budker Institute of Nuclear Physics SB RAS, 630090, Novosibirsk, Russia, [3]Novosibirsk State University, 630090, Novosibirsk, Russia.

Correspondence and requests for materials should be addressed to B.-F. S. (bfshen@mail.shcnc.ac.cn); or L.-L. J. (jill@siom.ac.cn).

**Nowadays, human's understanding of the fundamental physics is somehow limited by the energy that our high energy accelerators can afford. Up to 4 TeV protons are realized in the Large Hadron Collider (LHC). Leptons, such as electrons and positrons, however gained energies of about 100 GeV or less. Multi-TeV lepton accelerators are still lacking due to the relatively low acceleration gradient of conventional methods, which may induce unbearable cost. On the other hand, plasmas have shown extraordinary potential in accelerating electrons and ions, providing orders of magnitude higher acceleration fields of 10-100 GV/m. In such context, we propose a plasma-based**

**high-energy lepton accelerator, in which a weakly focusing plasma structure is formed near the beam axis. The structure preserves the emittance of the accelerated beam and produces low radiation losses. Moreover, the structure allows for a considerable decrease of the witness energy spread at the driver depletion stage.**

During the past few decades, vital researches on plasma-based accelerators have made great progress in generating high-energy electron beams. Nowadays, electrons with energies beyond 1 GeV are obtained experimentally[1-4]. In order to scale these approaches to the energy frontier of particle physics, i.e., to TeV energies, high energy proton beams, instead of laser pulses, were proposed to drive plasma wakefields and accelerate electrons[5], since they can bear much more energy than lasers drivers. A lepton collider for particle physics also requires positron beams of comparable energy and a good beam quality. However, it is simply impossible to apply the proposed blowout-like scenario[5,6] to positron acceleration. The main difficulty is that positrons, for their positive charges, are quickly expelled away from the bubble in the transverse direction[7-9] resulting in no efficient acceleration. There is no such a problem in the linear wakefield regime[10,11], but the linear regime has lower efficiency[12] and smaller accelerating fields, as compared to the nonlinear blowout regime.

To accelerate positrons by the plasma wakefield, the witness beam must be well confined in the accelerating region. We have found out that a simple hollow plasma channel can serve perfectly as the confiner[13,14]. The proposed scheme is sketched in Fig. 1. A cylindrical vacuum channel is embedded in the plasma. When a TeV proton beam travels through, it sucks plenty of electrons from the channel walls and drives a highly nonlinear plasma wave. The drawn away electrons generate a strong charge separation field at the boundary, which is directed to the propagating axis. In the formed structure, the positrons are both accelerated and transversely confined by the proton-driven wakefield. As will be shown later, it makes possible acceleration in a single stage of the length of about 1 kilometer, leading to the TeV energy, high quality positron beams.

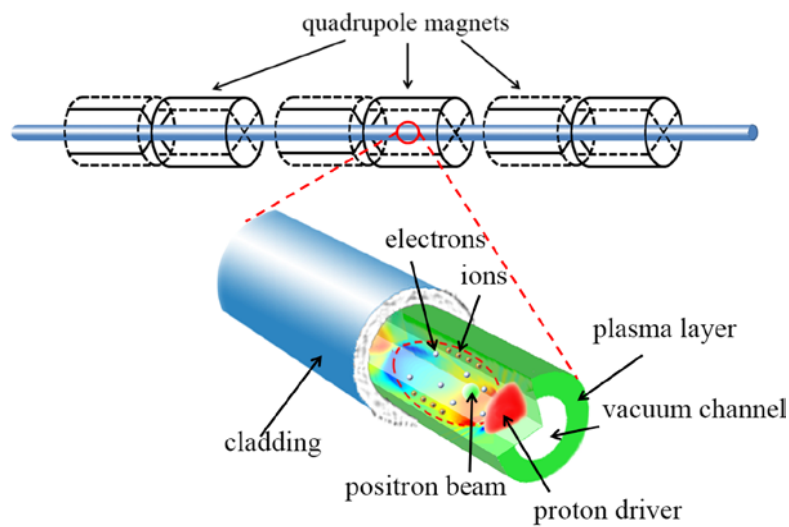

Fig. 1 (color online) Conceptual map of plasma wakefield acceleration in a hollow plasma channel. The quadrupole magnets are used to focus the proton driver. The accelerating field is shown by the colors.

The proposal shows several remarkable advantages. By optimizing the witness location, the total transverse force acting on the positrons almost vanishes. The beam then stays in a steady state without apparent transverse oscillations. Consequently, a very low transverse emittance can be maintained during the whole interaction length, and the radiation loss is tremendously suppressed.

## Results

We demonstrate the overall accelerating process with two-dimensional (2D) simulations. A 2 TeV proton beam, which energy is available in nowadays conventional accelerators, is used as the driver. In order to overcome natural disperse during the 1 km-propagation in plasmas, external quadrupole magnets are used to provide the required focusing as shown in Fig. 1. The quadrupole focusing of the beam is not fully consistent with the axisymmetric geometry of the problem, so we use the axisymmetric analogue of quadrupole fields described in Ref.[6]. An initially 1 GeV positron beam is injected behind the driver. The plasma has a constant initial density of $n_0 = 10^{15}$ cm$^{-3}$ outside the plasma channel $(r \geq r_0)$, where $r_0 = 0.75$ mm is the radius of the vacuum channel. Note that the sharp channel boundaries are not the essential part of the concept, so the channel will also work with diffuse edges as well. The channel itself is wider than state-of-the art channels produced for laser driven wakefield acceleration. These factors make these channels conceivable, though challenging due to the long length required. Detailed simulation parameters can be found in **Methods.**

**The hollow plasma channel enables a good radial confinement of the witness bunch.** The accelerating length as long as 1.1 km is realized in a single stage. Figure 2(a-d) shows snapshots of the particle phase space (energy vs distance) at several moments. It is clearly seen that during the propagation though the plasma channel, the proton driver is gradually stretched, losing significant amounts of energy at the tail.

Simultaneously, the witness positron beam is catching up with the driver and continuously picking up energies, until it reaches the dephasing region at around 1.1 km. The corresponding energy spectra are displayed in Figure 2(e-h), respectively. At the end of acceleration, a positron beam with the peak energy of about 1.6 TeV is produced.

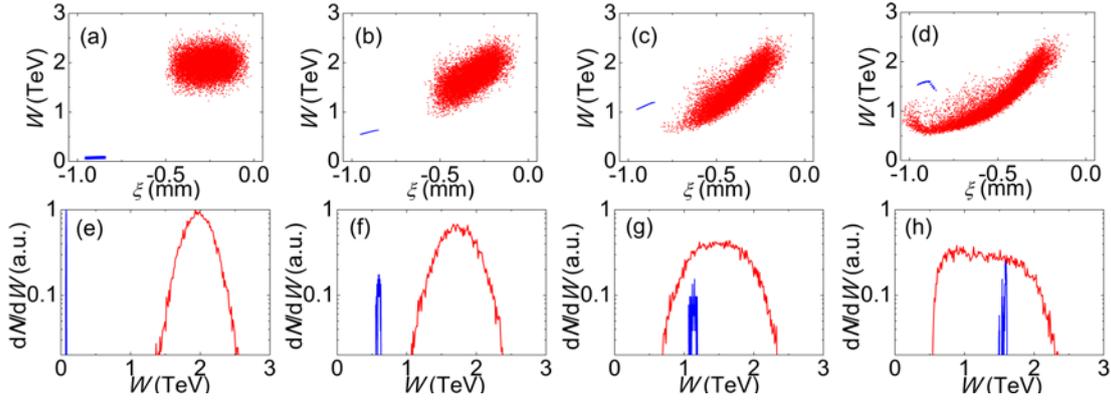

Fig. 2 (color online) Phase space portraits of the driver and the witness bunch (energy $W$ versus co-propagating coordinate $\xi = z - ct$) (a–d) and corresponding energy spectra (e–h). The snapshots are taken at propagation distances $L$ = 50, 400, 750, and 1100 m. The positrons and protons are shown in blue and red, respectively.

In Fig. 3(a), the structure of the proton-driven wakefield is shown. The accelerating field increases almost linearly in space along the upslope region (U region) and descends quickly in the downslope region (D region). As the positron beam enters the accelerating region, its peak energy grows proportional to the accelerating length, as is shown in Fig. 4(a). The energy gain saturates soon after the beam enters at the "D region" at around 900 meters.

**A novel compressing effect is seen in the behavior of the energy spread.** In Fig. 4(a), the energy spread stays at about 10% all along the "U region" and then remarkably drops in the "D region" to mere 1.5%. Though very narrow compared to "U region", the "D region" plays an important role in reducing the energy spread of the witness beam. More energetic positrons at the front experience weaker acceleration and vice versa, so that the longitudinal phase volume of the witness drastically shrinks, as presented in Fig. 2(d). We emphasize that this effect is specific to proton drivers, for which the driver-to-witness distance is continually reducing during the interaction. The positive charge of the witness is also important, since negatively charged electrons must reside on the left of the "U region" in Fig. 3(a), and the driver-to-witness distance must increase for the witness to cross the field extreme. Note also that the observed reduction of the energy spread does not rely on beam loading. Thus, the discovered "ballistic" reduction of the energy spread is independent and complimentary (to the beam loading) way of controlling the energy spread.

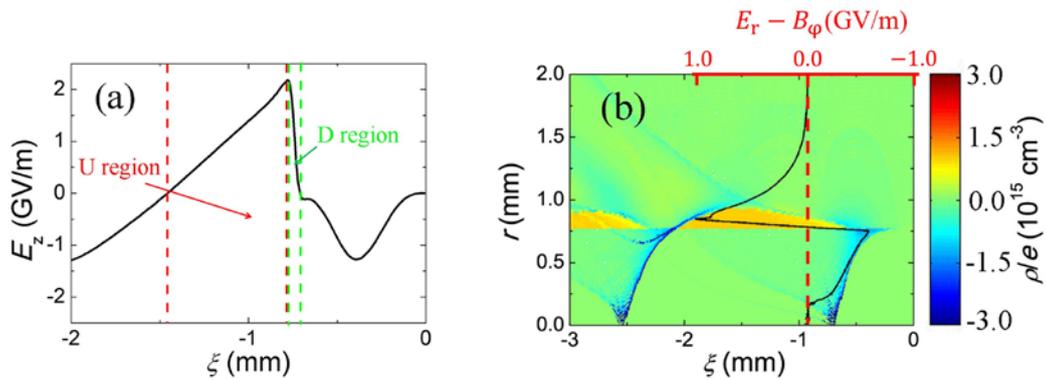

Fig. 3 (color online) (a) On-axis acceleration field of the plasma wave. (b) The net charge density distribution and the radial field $E_r - B_\varphi$ at the cross section of witness positrons (black solid line).

**A weakly focusing structure is formed near the beam axis as shown in Fig. 3(b), which produces low radiation losses and preserves the emittance of the**

**accelerated beam.** Generation of the high quality TeV positron beam in 1km implies not only good channeling and acceleration, but also suppression of the radiation loss. On the other hand, in order to build an electron-positron collider in future, a small beam emittance is required. The weakly focusing plasma structure in the proposed scheme gives a possible solution to these questions as shown in Fig. 4(b-d).

One can see that in the witness location, the transverse field $E_r - B_\varphi$ is almost-zero in the region close to the beam axis. The underlying physics for the formation of such structure is the introduction of hollow channel greatly weakens the "phase mixing" effect[15], thus reducing the electron density in the "bubble" (especially near the beam axis) as shown in Fig. 3(b). This is due to the plasma electrons are sucked into the hollow channel with an initially pitch angel (with respect to the beam aixs) larger than a certain value (determined by the channel radius).

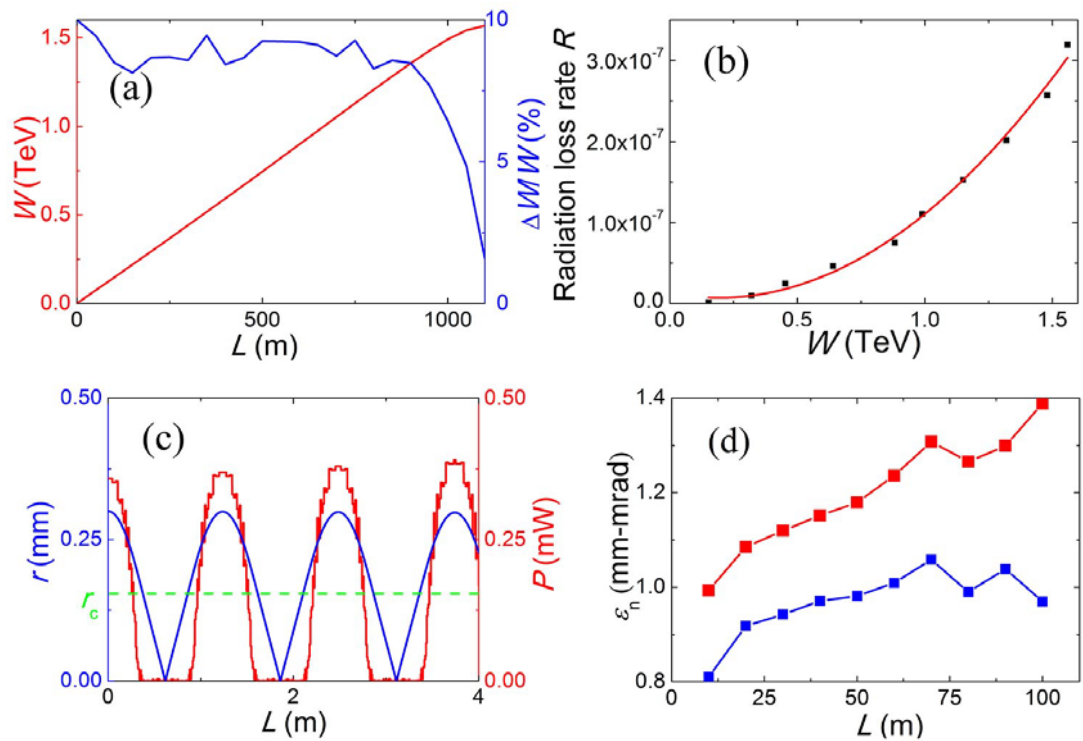

Fig. 4 (color online) (a) The mean energy (red) and energy spread (blue) of the witness positrons as a function of propagation distance. (b) Radiation loss rate (radiation loss / energy gain) for the positron beam as a function of witness energy (dots) and the $\gamma^2$ scaling (red line). (c) Time dependence of the radial position (blue) and radiation power (red) for a test positron initially at $r_{\text{in}} = 0.3$ mm, the green dashed line shows the critical radius $r_c$. (d) Normalized emittance of the full witness bunch (red) and core particles (blue) in first 100 meters of the acceleration.

A distinctive feature shows up near the axis in Fig. 4(b), where all fields vanish within a critical radius $r_c$. The beauty of such a structure is that the radiation of paraxial positrons (r ≤ $r_c$) is almost zero, resulting in a radiation-free region. To see this more clearly, the radiated power of a single test positron is plotted together with its trajectory in Fig. 4(c). The radiation power disappears when the positron enters the near axis region of $r \leq r_c \approx 0.15$ mm, just as we predicted. The abaxial positrons, or the ones with large pitch angles, which try to escape from the confining potential well, would be reflected by the strong focusing force, lose transverse velocity by radiation and then fall into the radiation-free region.

The weakly focusing plasma structure also allows for the conservation of the normalized emittance during the acceleration as shown by Fig. 4(d). A much finer grid is used here to follow the emittance growth, so that we only simulate for the first 100 meters in the acceleration due to the limitations of the computing time. One can see that the normalized emittance of the full bunch grows slightly. A few halo particles are responsible for this emittance growth (as discussed in the **Discussion**). If those particles are excluded, the emittance is almost conserved for the core particles

(containing 95% of the total charge). As we use a quasi-static code, there is no numerical Cherenkov radiation inherent to particle-in-cell codes[16,17], the unphysical emittance growth is fully controlled by grid and time steps.

## Discussion

The radiation loss may be important for leptons accelerated to TeV range energies. Previous studies[11] indicate the betatron oscillation produces significant energy damping for the electron accelerated in linear plasma wave. Positrons at TeV-level have relativistic factors of the order of $10^6$. Hence the radiation loss in the channel mainly comes from their transverse oscillations.

Assuming the witness bunch propagates towards positive z-direction, the radiation power for a positron can be calculated as[18]

$$P \approx \frac{2r_e^2 c}{3e^2}\gamma^2\{(E_r - B_\varphi)e + f_r + f_q\}^2, \qquad (1)$$

where $r_e = e^2/mc^2$ is the classic electron radius, $e$ and $m$ are the positron charge and mass, $\gamma = (1 - v^2/c^2)^{-1/2}$ is the relativistic factor, and $c$ is the light velocity, respectively; $E_r$ and $B_\varphi$ are the radial electric field and azimuthal magnetic field in the channel. The term $f_q$ is the time-average force of the quadrupole magnets given by[6]

$$f_q = -\frac{S^2 L_q^2 e^2 r}{8\pi^2 W_w}, \qquad (2)$$

where $S$ is the magnetic field gradient of the quadrupoles, $L_q$ is the space period, and $W_w$ is the energy of witness bunch. The term $f_r$ in Eq. (1) is the radial component of the radiation reaction (RR) force on the positron, which, in the relativistic case, is[19]

$$f_\mathrm{r} \approx -\frac{2r_e^2}{3}\beta_r\left\{[E_z^2(1+\gamma^2\theta^2)]+\gamma^2(E_\mathrm{r}-B_\varphi)^2\right\}, \quad (3)$$

with $E_\mathrm{z}$ the accelerating field, $\theta = v_r/v_\mathrm{z}$ the angle between the positron velocity and the axis, and $\beta_r$ the transverse velocity of the positron in units of $c$. Since $\theta \ll 1$, the items related to $E_\mathrm{r} - B_\varphi$ are the main source of the radiation loss. The radial Lorentz force $E_\mathrm{r} - B_\varphi$ is shown in Fig. 3(b). Inside the channel ($r \leq r_0$), the focusing field is created by plasma electrons pulled out of the boundary.

One can numerically calculate the radiation loss rate from Eqs. (1-3). Let $G = eE_z v_z$ be the energy gain of the witness positron per unit time. We define the radiation loss rate by $R = P/G$. A bunch of test positrons is tracked in the electro-magnetic fields taken from simulations. As predicted by Eq. (1), the quadratic dependence of the radiation loss rate on the witness energy holds up to multi-TeV regime [Fig. 4(b)]. In our case, $R$ is only $3.2 \times 10^{-7}$ for 1.6 TeV energy. The radiation loss is extremely low so that more precise calculations are unnecessary.

Since the radial force below the critical radius is almost zero, the positrons accelerated there are free from emittance growth[20,21]. Nonetheless, there would always be some positrons escaping from this area and being reflected by the boundary field. As the threshold $r_\mathrm{c}$ varies slightly with time (acceleration distance), such reflection is usually not a "mirror reflection", resulting in increase of the beam emittance. It is thus desirable to avoid touching the critical radius by the witness body. Let us estimate the witness emittance required for that. Considering a test positron initially located at radius $\sigma_\mathrm{wr}$, having pitch angle $\alpha_\mathrm{i}$ and the relativistic factor $\gamma_\mathrm{i}$. This positron can be thought as a marginal one for the positron beam of the normalized emittance

$\varepsilon_{cn} \approx \gamma_i \alpha_i \sigma_{wr}$. If uniformly accelerated by the field $E_z$ to the relativistic factor $\gamma_f$, this positron will change its radial position to

$$r_f = \sigma_{wr} + \int_0^L \gamma_i \alpha_i / \gamma(z) dz = \sigma_{wr} + \gamma_i \alpha_i \ln\left(\frac{\gamma_f}{\gamma_i}\right) \frac{mc^2}{eE_z}. \tag{4}$$

Assuming $r_f = r_c$, we find the threshold witness emittance

$$\varepsilon_{cn} = \frac{eE_z}{mc^2} \left[\ln\left(\frac{\gamma_f}{\gamma_i}\right)\right]^{-1} (r_c \sigma_{wr} - \sigma_{wr}^2). \tag{5}$$

If the emittance is below this value, the most part of the witness propagates in vacuum and sees no particles or transversely deflecting fields on its path.

For the present parameters, Eq. (5) gives the threshold of approximately 2 mm-mrad. We then chose the initial witness emittance of 1 mm-mrad in our simulation. The normalized emittance almost conserves in the simulation as displayed in Fig. 4(d), which confirms our discussion above.

In conclusion, we have suggested a PWFA scheme capable of accelerating positrons to energies beyond 1 TeV. The novel feature of this scheme is a combination of the proton driver and the vacuum channel created inside a uniform plasma. An initially 1 GeV positron beam is not only accelerated efficiently by the proton-driven wakefield, but also confined radially in the almost rectangular potential well. In this scheme, the radiation damping is negligibly small, and the normalized emittance is almost preserved for the core particles. The relatively low velocity of the proton driver is turned to advantage and used to reduce the final energy spread of accelerated positrons. In the illustrative simulation, a quasi-monoenergetic positron beam is obtained in a single acceleration stage, with the final energy of 1.6 TeV and the energy spread of only 1.5%.

## Methods

The simulations are performed with computationally efficient code LCODE[22] capable of simulating long distance beam propagation. The simulation is carried out in the axisymmetric geometry. In the code, the simulation window moves with the light velocity. The beams and plasma are modeled by fully relativistic macro-particles. Quasi-static approximation is used for calculating the plasma response so that the radiation reaction is not included in this code and was estimated separately. Two different simulation grid sizes are used in this work. The finer one of size $0.01c/\omega_p \approx 1.68$ μm is used to study the detailed shape of the plasma wave and the emittance conservation ($\omega_p$ is the plasma frequency of the surrounding plasma), and the coarse one of the size $0.05c/\omega_p \approx 8.4$ μm is used to simulate acceleration process over the long distance, The time steps for these two cases are $5\omega_p^{-1} \approx 2.8$ ps and $12\omega_p^{-1} \approx 6.7$ ps, respectively.

The detailed simulation parameters are listed in Table 1. The parameters of the driver are similar to those used in Refs. [5] and [13]. The driver energy is larger (2 TeV) so that it can accelerate positrons beyond 1 TeV, and a typical 1 GeV positron beam serves as the witness. We have scanned over a range of plasma densities (near the matching condition $k_p\sigma_z \sim \sqrt{2}$[5,9]) and channel radii and found out that $n_0 = 10^{15}$ cm$^{-3}$ and $r_0 = 0.75$ mm provides the best performance.

| Parameters | Symbols | Values |
| --- | --- | --- |
| Number of protons in the drive beam | $N_p$ | $10^{11}$ |
| Initial energy of protons | $W_p$ | 2 TeV |
| Initial longitudinal size of the driver | $\sigma_z$ | 100 μm |
| Initial spot size of the driver | $\sigma_r$ | 430 μm |
| Initial longitudinal momentum spread of the driver | $\Delta P_z/P_z$ | 0.1 |
| Normalized emittance of the driver | $\varepsilon_n$ | 1 mm-mrad |
| Initial witness energy | $W_w$ | 1 GeV |
| Total witness charge | $Q_w$ | 800 pC |
| Initial bunch length of the witness beam | $\sigma_{wz}$ | 25 μm |
| Initial spot size of the witness beam | $\sigma_{wr}$ | 50 μm |
| Initial normalized emittance of the witness beam | $\varepsilon_{wn}$ | 1 mm-mrad |
| Plasma density | $n_0$ | $10^{15}$ cm$^{-3}$ |
| Plasma channel radius | $r_0$ | 0.75 mm |
| Magnetic field gradient | $S$ | 0.5 T/mm |
| Space period of the quadrupoles | $L_q$ | 3 m |

Table 1 Parameters for the simulation.

# References

1	Tajima T. & Dawson J. M. Laser electron accelerator. *Phys. Rev. Lett.* **43**, 267-270 (1979)

2	Leemans, W. P. et al. GeV electron beams from a centimetre-scale accelerator. *Nat. Phys.* **2**, 696-699 (2006).

3	Wang, X. *et al.* Petawatt-laser-driven wakefield acceleration of electrons to 2 GeV in $10^{17}$ cm−3 plasma. in *Proceedings of AIP Conference,* **1507** 341-344, doi: 10.1063/1.4773719 (2012).

4	Lu, H. et al. Laser wakefield acceleration of electron beams beyond 1 GeV from an ablative capillary discharge waveguide. *Appl. Phys. Lett.* **99**, 091502 (2011).

5	Caldwell, A., Lotov, K., Pukhov, A. & Simon, F. Proton-driven plasma-wakefield acceleration. *Nat. Phys*. **5**, 363-367 (2009).

6	Lotov, K. V. Simulation of proton driven plasma wakefield acceleration. *Phys. Rev. Spec. Top.- Accel. Beams* **13**, 041301 (2010).

7	Lotov, K. V. Acceleration of positrons by electron beam-driven wakefields in a plasma. *Phys. Plasmas* **14**, 023101 (2007).

8	Hogan, M. J. *et al.* Plasma wakefield acceleration experiments at FACET. *New J. Phys.* **12**, 055030 (2010).

9	Yi, L. *et al.* Proton acceleration by plasma wakefield driven by an intense proton beam. *Laser Part. Beams*. **31**, 427-438, (2013)

10	Nakajima, K. *et al.* Operating plasma density issues on large-scale laser-plasma accelerators toward high-energy frontier. *Phys. Rev. Spec. Top.- Accel. Beams* **14**, 091301 (2011).

11   Deng, A. *et al.* Electron beam dynamics and self-cooling up to PeV level due to betatron radiation in plasma-based accelerators. *Phys. Rev. Spec. Top.- Accel. Beams* **15**, 081303 (2012).

12   Lotov, K. V. Efficient operating mode of the plasma wakefield accelerator. *Phys. Plasmas* **12**, 1889444 (2005).

13   Yi, L. *et al.* Scheme for proton-driven plasma-wakefield acceleration of positively charged particles in a hollow plasma channel. *Phys. Rev. Spec. Top.- Accel. Beams* **16**, 071301 (2013).

14   Schroeder C. B., Esarey E., Benedetti C., & Leemans W. P. Control of focusing forces and emittances in plasma-based accelerators using near-hollow plasma channels. *Phys. Plasma* **20**, 080701 (2013).

15   Kimura, W. D., Milchberg, H. M., Muggli, P., Li, X. & Mori, W. B. Hollow plasma channel for positron plasma wakefield acceleration. *Phys. Rev. Spec. Top.- Accel. Beams* **14**, 14.041301 (2011).

16   Lehe, R., *et.al*., Numerical growth of emittance in simulation of laser-wakefield acceleration. *Phys. Rev. Lett.* **16**,021301 (2013).

17   Huang, C. *et al.*, QUICKPIC: A highly efficient particle-in-cell code for modeling wakefield acceleration in plasmas. *J. Comput. Phys.* **217**, 658, (2006).

18   Jackson J. D., [Chapter 14 Radiation by Moving Charges] *Classical*

*Electrodynamics, 3rd ed.* [665-667] (Wiley, New York, 1999).

19   Landau, L. D., & Lifshitz E. M., [Chapter 9 Radiation of Electromagnetic Waves] *The Classical Theory of Fields, 4th ed.* [226-229] (Pergamon, New York, 1994).

20   Assmann, R. & Yokoya, K., Transverse beam dynamics in plasma-based linacs. *Nucl. Instru. Methods Phys. Res. Sect. A:* **410**, 544-548, (1998).

21   Brunetti, E. *et al.* Low Emittance, High Brilliance Relativistic Electron Beams from a Laser-Plasma Accelerator. *Phys. Rev. Lett.* 105, 215007 (2010).

22   Lotov, K. Fine wakefield structure in the blowout regime of plasma wakefield accelerators. *Phys. Rev. Spec. Top.- Accel. Beams* **6**, 061301 (2003).

**Acknowledgements**

This work has been supported by the Ministry of Science and Technology (2011CB808104, 2011DFA11300), National Natural Science Foundation of China (No. 11125526, No. 11335013, No. 11374317, No. 11127901, and No. 60921004), and by the Ministry of Education and Science of the Russian Federation.

**Author contributions**

L-Q. Y., B-F. S., and L-L. J. contributed to all aspects of this work; K. L. and A. S. contributed to code development; X-M. Z., W-P. W., J-C. X. and K. L. provided inspiring ideas and helped L-Q. Y. to write the paper; Y. S. and L-G. Z. helped to analyze the radiation loss in the acceleration process; Z-Z. X. gave some useful

## Additional information

Competing financial interests: The authors declare no competing financial interests.

The code LCODE used in this work is freely available online at http://www.inp.nsk.su/~lotov/lcode/